\documentclass[twocolumn,showpacs,amsmath,amssymb]{revtex4}
  
\usepackage{graphicx}
\usepackage{dcolumn}
\usepackage{bm}
\usepackage{bbm}
\usepackage{amsfonts}

\def\>{\rangle}
\def\<{\langle}
\def\ket#1{|#1\>}
\def\bra#1{\<#1|}
\def\braket#1#2{\< #1 | #2 \>}

\def\ave#1{\left\< #1\right\>}

\def\ii{{\rm i}}
\def\dd{{\rm d}}
\def\TT{{\cal\hat T}}

\begin{document}

\title{Quantum freeze of fidelity decay for chaotic dynamics}
\author{Toma\v z Prosen and Marko \v Znidari\v c}
\affiliation{Physics Department, Faculty of Mathematics and Physics, 
University of Ljubljana, Ljubljana, Slovenia}

\date{\today}

\begin{abstract}
We show that the mechanism of {\em quantum freeze} of fidelity decay for
perturbations with zero time-average, recently discovered for 
a specific case of integrable dynamics [New J. Phys. {\bf 5} (2003) 109], 
can be generalized to arbitrary quantum dynamics. We work out explicitly 
the case of chaotic classical counterpart, for which we find semi-classical 
expressions for the value and the range of the plateau of 
fidelity. After the plateau ends, we find explicit expressions for the
asymptotic decay, which can be exponential or Gaussian depending on the 
ratio of the Heisenberg time to the decay time. Arbitrary 
initial states can be considered, e.g. we discuss coherent states and 
random states.
\end{abstract}
\pacs{03.65.Yz, 03.65.Sq, 05.45.Mt}

\maketitle

The question of stability of quantum time evolution with respect to small
changes in the Hamiltonian has recently attracted lot of attention 
\cite{fidelity,jalabert}. This question is particularly important in the context
of quantum information processing \cite{qcomp}. The central quantity for 
describing quantum stability is the fidelity 
$F(t) = |\braket{\psi(t)}{\psi_\delta(t)}|^2$ where 
$\ket{\psi(t)}=U_0(t)  \ket{\psi}$ and 
$\ket{\psi_\delta(t)}=U_\delta(t) \ket{\psi}$ are unperturbed and
perturbed time evolutions, of perturbation strength $\delta$, respectively, 
starting from the same {\em initial state} $\ket{\psi}$.
Let the evolution operator be written as time-ordered product 
$U_\delta(t)= \TT \exp(-\ii\int_0^t \dd t' H_\delta(t')/\hbar)$
in terms of (generally time-dependent) 
Hamiltonian $H_\delta(t)=H_0(t) + \delta H'(t)$. In this Letter we either 
assume that $H_\delta(t)$ is autonomous (time-independent), or more generally,
periodically  time dependent with some period $\tau$, $H_\delta(t+\tau) = H_\delta(t)$.
Then the time is measured in discrete units of $\tau$, namely $t=n\tau$,
and the former (autonomous) case is simply obtained as the limit $\tau\to 0$.
The perturbed propagator for one time step can be written as
$U_\delta(\tau) = U_0(\tau)\exp(-\ii V \tau\delta/\hbar)$ in terms of a hermitean
perturbation $V$ which in the leading order perturbs the Hamiltonian
$V = H' + {\cal O}(\tau \delta)$.

It has been shown \cite{jalabert} that for classically chaotic systems and
for sufficiently strong perturbation and coherent initial state 
$\ket{\psi}$ the fidelity decay is given by classical Lyapunov exponents, 
and this phenomenon has been recently explained solely on the basis of 
classical dynamics \cite{VP}. On the other hand, for sufficiently 
small $\delta$, one can express fidelity decay in terms of a power series 
in $\delta$ where coefficients are given as time-correlation function of 
the perturbation \cite{PZ}. Using this approach one can derive universal
forms of fidelity decay in both cases of classically regular and chaotic 
dynamics and express all time-scales solely in terms of classical
quantities and $\hbar$.

The starting point of our analysis is the representation of fidelity $F^{(n)} = F(n\tau)$ 
in terms of expectation value \cite{PZ} 
\begin{equation}
F^{(n)} = |f^{(n)}|^2,\quad f^{(n)} = \bra{\psi}M_\delta^{(n)}\ket{\psi} 
\label{eq:fid}
\end{equation} 
of the {\em echo operator} 
$M_\delta^{(n)}:=U_0(-n\tau) U_\delta(n\tau)$ 
which is the propagator in the interaction picture. Namely 
\begin{equation}
M_\delta^{(n)} = \TT\exp\left(-\ii\frac{\delta}{\hbar}
\Sigma_n(V)\right),
\label{eq:echoop}
\end{equation}
where, for any operator $A$, $\Sigma_n(A):=\tau \sum_{n'=0}^{n-1} A_{n'}$ and
$A_n:=U_0(-n\tau)A U_0(n\tau)$.
In case of continuous time: $M_\delta(t)=\TT\exp(-\ii\frac{\delta}{\hbar}\Sigma(V,t))$,
with $\Sigma(A,t) := \int_0^t \dd t' A(t')$, $A(t):=U_0(-t)A U_0(t)$. 
Approach \cite{PZ} using the power law 
expansion of (\ref{eq:echoop}) in $\delta$ gives to the second order
\begin{equation}
F^{(n)} = 1 - \frac{\delta^2}{\hbar^2}\left\{ \ave{\Sigma^2_n(V)} - \ave{\Sigma_n(V)}^2 \right\}
+ {\cal O}(\delta^4)
\end{equation}
where $\ave{\bullet}:=\bra{\psi}\bullet\ket{\psi}$. Equivalently, this useful formula can be 
expressed in terms of time-correlation function 
$C(n',n'') = \ave{V_{n'} V_{n''}} - \ave{V_{n'}}\ave{V_{n''}}$, namely
$F^{(n)} = 1-(\tau \delta/\hbar)^2\sum_{n',n''=0}^{n-1} C(n',n'') + \ldots$
The rule of thumb says that slower decay of correlations, {\em i.e.} 
stronger fluctuation of $\Sigma_n(V)$, imply faster decay of fidelity, 
and vice versa.

Particularly interesting special situation arises when a
time averaged perturbation $\bar{V}:=\lim_{n\to\infty}(n\tau)^{-1}\Sigma_n(V)$
equals zero. In general, the perturbation can be decomposed into the {\em diagonal} 
and {\em residual} part $V = \bar{V} + V_{\rm res}$. The part 
$\bar{V}$ which commutes with the unperturbed evolution $U_0$ and is thus 
{\em diagonal} in its eigenbasis can sometimes be put together with the 
unperturbed Hamiltonian $H_0$. This is customary in various quantum mean 
field approaches. It is thus interesting
to question the stability of quantum dynamics with respect to residual 
perturbation only (i.e. when its diagonal part exactly vanishes $\bar{V}=0$).
This problem has been addressed for the particular case of perturbed
integrable dynamics \cite{ifreeze} and very interesting results on extreme stability of 
quantum dynamics have been found termed as 'quantum freeze of fidelity'. 

In this Letter we show that the
phenomenon of quantum freeze, namely the saturation of fidelity to a plateau of high value, 
is much more general and robust as it appears in Ref.\cite{ifreeze}, 
and applies to arbitrary quantum evolution provided only that the 
perturbation is residual, $\bar{V}=0$. In particular, we work out in detail 
the important case of dynamics with fully chaotic classical counterpart.
We compute the plateau value (scaling as $1-{\rm const} \delta^2$ within the
second order), its range scaling as $1/\delta$, and the rate of 
the asymptotic decay after the plateau ends (which is either gaussian or exponential), 
quantitatively in terms of the underlying classical dynamics, effective perturbation $\delta$
and the effective value of Planck constant $\hbar$.
The phenomenon may find useful application in quantum 
computation where fidelity error is predicted to be very small and frozen in time 
(for sufficiently small $\delta$) provided only that the diagonal
part of the error in each gate can be cured by some other means.

In the autonomous case ($\tau\to 0$), provided that the spectrum of $H_0$ is non-degenerate
(which is true for a generic non-integrable system), the perturbation is
residual iff it can be written as a {\em time derivative} of some observable $W$
{\em i.e.} a commutator with $H_0$, $V = \frac{\ii}{\hbar}[H_0,W] = (d/dt)W$. 
Generalizing to the discrete, time-periodic case we shall assume that the perturbation is of the 
form
\begin{equation}
V = \frac{1}{\tau}\left(W_1 - W_0\right) = \frac{1}{\tau}\left(U_0(-\tau)W U_0(\tau) - W\right).
\label{eq:pert}
\end{equation}
We shall now apply the Baker-Campbell-Hausdorff expansion 
$e^A e^B = \exp(A+B+(1/2)[A,B]+\ldots)$ to the time-ordered product (\ref{eq:echoop}) and
rewrite the echo-operator
\begin{equation}
M^{(n)}_\delta = \exp\left\{ -\frac{\ii}{\hbar}\left(\Sigma_n(V) \delta 
+ \frac{1}{2} \Gamma_n \delta^2 + \ldots\right)\right\}
\label{eq:BCH}
\end{equation}
where 
$\Gamma_n:= \frac{\ii\tau^2}{\hbar}\sum_{n'=0}^{n-1}\sum_{n''=n'}^{n-1} [V_{n'},V_{n''}]$.
It is interesting to note \cite{ifreeze} that all matrix elements of $\Gamma_n$ grow with $n$
{\em not faster} than $\propto n$ \cite{control}. 
This becomes obvious for the special form of perturbation 
(\ref{eq:pert}) for which it follows
\begin{eqnarray}
\Sigma_n(V) &=& W_n - W_0,\\
\Gamma_n &=& \Sigma_n(R)-\frac{\ii}{\hbar}[W_0,W_n],
\quad R := \frac{\ii}{\tau\hbar}[W_0,W_1],\qquad
\label{eq:defR}
\end{eqnarray}
so the operator $\Gamma_n$ is also a time sum/integral
of a time-dependent operator $R$, minus a sort of time-correlation function which shall be neglected
for systems with strong decay of correlations studied below.
In the continuous time case, $R = \frac{\ii}{\hbar}[W,(d/dt)W] = \hbar^{-2}[W,[W,H_0]]$ and
$\Gamma(t) = \int_0^t \dd t' R(t') - \frac{\ii}{\hbar}[W(0),W(t)]$.
We note that, provided $W$ has a well defined classical limit $\hbar\to 0$, 
then also $V$, $R$ and $\Gamma_n$ have well defined limits since 
$\frac{\ii}{\hbar}[\bullet,\bullet]$ can
be replaced by a Poisson bracket. This is what we shall assume below, as well as that 
the limiting classical dynamics of $U_0$ is fully chaotic.

Comparing the two terms in the BCH exponential (\ref{eq:BCH}) we note that there should 
exist a time-scale $t_2 \sim \delta^{-1}$, such that if $n\tau < t_2$ then the 
first term $\Sigma_n \delta$ dominates the second one $\frac{1}{2}\Gamma_n \delta^2$ 
(and higher \cite{control}).
So, let us first consider the case $n \tau < t_2$.
Then we can neglect the second term and write the fidelity amplitude
\begin{equation}
f^{(n)} = \ave{\exp\left(-\ii(W_n - W_0)\delta/\hbar\right)}.
\label{eq:fshort}
\end{equation}
Expanding $f^{(n)}$ to the second order in $\delta$, we find
$F^{(n)} = 1 - \frac{\delta^2}{\hbar^2}(\kappa_0^2 + \kappa_n^2 - C_n - C_n^*)$ where
$\kappa_k^2:=\ave{W_k^2}-\ave{W_k}^2$, $C_n:=\ave{W_n W_0}-\ave{W_n}\ave{W_0}$.
Using Cauchy-Schwartz inequality $|C_n| \le \kappa_0 \kappa_n$ and the fact that for a bounded 
operator $W$ the sequence $\kappa_n$ is bounded, say by $r$, we find a freeze of fidelity
$1-F^{(n)} \le 4\frac{\delta^2}{\hbar^2}r^2$, $n\tau < t_2\propto\delta^{-1}$, 
for {\em arbitrary} quantum dynamics, irrespective of the existence and the nature of the 
classical limit. 

Let us further assume that, due to mixing property of classically chaotic dynamics,
time correlations vanish semiclassically beyond some {\em mixing time-scale} $t_1$, 
$C_n \to {\cal O}(\hbar)$ for $n\tau > t_1$, and quantum expectation values become 
time-independent and equal, to leading order in $\hbar$,
 to the classical averages over an appropriate invariant set 
$\ave{A}_{\rm cl} := \int\dd\mu A_{\rm cl}$. 
Hence, between $t_1$ and $t_2$, 
the fidelity freezes to a constant value \cite{appr}
\begin{equation}
F_{\rm plat} \approx 1 - \frac{\delta^2}{\hbar^2}\left(\kappa^2_0 + \kappa^2_{\rm cl}\right), \quad
\kappa^2_{\rm cl} := \ave{W^2}_{\rm cl} - \ave{W}_{\rm cl}^2.
\label{eq:LRP}
\end{equation}
Considering two interesting extreme examples of initial states, namely {\em coherent initial states} (CIS)
and {\em random initial states} (RIS) we find: For CIS $\kappa^2_0\propto \hbar$ can be neglected with 
respect to $\kappa^2_{\rm cl}$, whereas for RIS $\kappa^2_n$ does not depend on time hence 
$\kappa^2_0 = \kappa^2_{\rm cl}$. So within the linear response approximation $1-F_{\rm plat}$ is 
universally twice as large for RIS than for CIS. It is also worth to stress that quantum relaxation time 
$t_1 \propto \log\hbar$ for CIS while $t_1 \propto \hbar^0$ is simply the classical mixing time for RIS.

One can go beyond the linear response in approximating (\ref{eq:fshort}) using a simple fact that
in the leading order in $\hbar$ quantum observables commute, and as before, that for $n\tau > t_1$ the 
time correlations vanish, namely $\ave{\exp(-\ii\frac{\delta}{\hbar}(W_n-W_0))} \approx
\ave{\exp(-\ii\frac{\delta}{\hbar}W_n)}\ave{\exp(\ii\frac{\delta}{\hbar} W_0)}$:
\begin{equation}
F_{\rm plat} \approx 
\left|\ave{\exp{(-\ii W \delta/\hbar)}}_{\rm cl}\ave{\exp{(\ii W_0\delta/\hbar)}}\right|^2.
\label{eq:NLRP}
\end{equation}
Defining a generating function in terms of the classical observable $W_{\rm cl}$, 
$G(z):=\ave{\exp(-\ii z W)}_{\rm cl}$, one can compactly write
$F^{\rm CIS}_{\rm plat}\approx|G(\delta/\hbar)|^2$ for CIS (neglecting localized initial state average 
with $W_0$) and
$F^{\rm RIS}_{\rm plat}\approx |G(\delta/\hbar)|^4$ for RIS, satisfying {\em universal} relation
$F^{\rm RIS}_{\rm plat}\approx (F^{\rm CIS}_{\rm plat})^2$. Curiously, the same relation is satisfied for the
case of regular dynamics \cite{ifreeze}.
If the argument $z=\delta/\hbar$ is large, the analytic function $G(z)$ can be 
calculated generally by the method of stationary phase.
In the simplest case of a single isolated stationary point $\vec{x}^*$ in $N$ dimensions:
\begin{equation}
|G(z)| \asymp \left|\frac{\pi}{2z}\right|^{N/2}\left|{\rm det\,}\partial_{x_j} \partial_{x_k} 
W(\vec{x}^*)\right|^{-1/2}.
\label{eq:Gasym}
\end{equation}
This expression gives an asymptotic power law decay of the plateau height 
independent of the perturbation details. Note that for a finite phase space we will have 
oscillatory {\em diffraction corrections} to eq.~(\ref{eq:Gasym}) due to a finite range of 
integration $\int\dd\mu$ which in turn causes an interesting situation for 
specific values of $z$, namely that by 
increasing the perturbation strength $\delta$ we can actually increase the value of the plateau. 

Next we shall consider the regime of long times $n \tau > t_2$. 
Then the second term in the exponential of 
(\ref{eq:BCH}) dominates the first one, however even the first term may not be negligible.
Up to terms of order ${\cal O}(n \delta^3)$ we can factorize eq. (\ref{eq:BCH}) as
$
M_\delta^{(n)} \approx \exp(-\ii\frac{\delta}{\hbar}(W_n-W_0))
\exp(-\ii\frac{\delta^2}{2\hbar}\Gamma_n).
$
When computing the expectation value $f^{(n)} = \langle M_\delta^{(n)}\rangle$
we again use the fact that in the leading semiclassical order the operator ordering is irrelevant and
that, since $n\tau \gg t_1$, any time-correlation can be factorized,
so also the second term of $\Gamma_n$ (\ref{eq:defR}) vanishes. Thus we have
\begin{equation}
F^{(n)} \approx F_{\rm plat} \left|\ave{\exp\left(-\ii\frac{\delta^2}{2\hbar}\Sigma_n(R)\right)}\right|^2,
\quad n\tau > t_2.
\label{eq:renfid}
\end{equation}
This result is quite intriguing. It tells us that apart from a pre-factor $F_{\rm plat}$, the
decay of fidelity with residual perturbation is formally the same as fidelity decay with
a generic non-residual perturbation, eqs. (\ref{eq:fid},\ref{eq:echoop}), when one substitutes the 
operator $V$ with $R$ and the perturbation strength $\delta$ with $\delta_R=\delta^2/2$. 
The fact that time-ordering is absent in eq. (\ref{eq:renfid}) as compared with 
(\ref{eq:echoop}) is semiclassically
irrelevant. Thus we can directly apply the general semiclassical theory of fidelity decay \cite{PZ},
using a renormalized perturbation $R$ of renormalized strength $\delta_R$. 
Here we simply rewrite the key results in the 'non-Lyapunov' perturbation-dependent 
regime, $\delta_R\tau < \hbar$. Using a classical transport rate 
$\sigma := \lim_{n\to\infty}\frac{1}{2 n\tau}(\ave{\Sigma^2_n(R)}_{\rm cl}-\ave{\Sigma_n(R)}_{\rm cl}^2)$
we have either an exponential decay 
\begin{equation}
F^{(n)} \approx F_{\rm plat}
\exp{\left(-\frac{\delta^4}{2\hbar^2} \sigma n\tau\right)},\quad
n\tau < t_{\rm H}
\label{eq:Fexp}
\end{equation}
or a (perturbative) Gaussian decay 
\begin{equation}
F^{(n)} \approx F_{\rm plat}
\exp{\left(-\frac{\delta^4}{2\hbar^2} \sigma \frac{(n\tau)^2}{t_{\rm H}}\right)},\quad
n\tau > t_{\rm H}.
\label{eq:Fgau}
\end{equation}
$t_{\rm H}=\tau {\cal N}/(2s)$ is the Heisenberg time, where ${\cal N}\sim \hbar^{-d}$ 
(in $d$ degrees of freedom) is the
total dimension of the Hilbert space supporting the time evolution and $s$ is the
number of different symmetry classes (of possible discrete symmetries of $H_0$)
carrying the initial state $\ket{\psi}$.
This is just the time when the integrated correlation function of $R_n$ becomes dominated by quantum fluctuation.
Comparing the two factors in (\ref{eq:Fexp},\ref{eq:Fgau}), i.e. the fluctuations of two terms in
(\ref{eq:BCH}), we obtain a semiclassical estimate of $t_2$
\begin{equation}
t_2 \approx {\rm min}\left\{\sqrt{\frac{t_H}{\sigma}}\frac{\kappa_{\rm cl}}{\delta},
\frac{\kappa^2_{\rm cl}}{\sigma \delta^2}\right\}.\label{eq:t2}
\end{equation}
Interestingly, the exponential regime (\ref{eq:Fexp})
can only take place if $t_2 < t_{\rm H}$. 
If one wants to keep $F_{\rm plat} \sim 1$, or have exponential decay in the full range until 
$F \sim 1/{\cal N}$, this implies a condition on dimensionality: $d \ge 2$. 
The quantum fidelity and its plateau values have been
expressed (in the leading order in $\hbar$) in terms of classical quantities only.
While the prefactor $F_{\rm plat}$ depends on the details of initial state, 
the exp-factors of (\ref{eq:Fexp},\ref{eq:Fgau}) {\em do not}.
Yet, the freezing of fidelity is a purely quantum phenomenon. 
The corresponding {\em classical fidelity} (defined in \cite{PZ}) does not exhibit 
freezing. Let us now demonstrate our theory by numerical examples.
\par
\begin{figure}
\centerline{\includegraphics[angle=-90,width=3.2in]{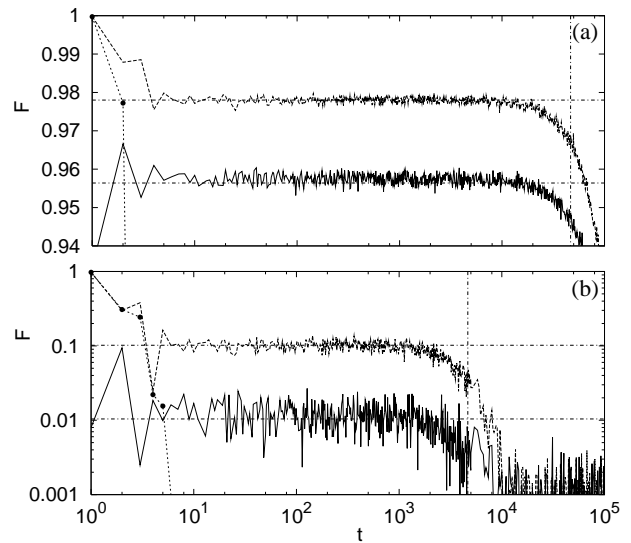}}
\caption{$F(t)$ for the kicked top, with $\delta=10^{-3}$ (a), and $\delta=10^{-2}$ (b). 
In each plot the upper curve is for CIS and the lower for RIS. Horizontal chain lines are 
theoretical plateau values (\ref{eq:NLRP}), vertical chain lines are theoretical values
of $t_2$ (\ref{eq:t2}). The full circles represent calculation of
the corresponding classical fidelity for CIS which follows quantum fidelity 
up to the Ehrenfest ($\log\hbar$) barrier and exhibits no freezing.}
\label{fig:1ktop}
\end{figure}
First we consider a quantized kicked top as an example of one-dimensional system ($d=1$).
The system is described by quantum angular momentum $J_{\rm x,y,z}$ with (half)integer
modulus $J$ and the one-step propagator
$U=\exp{(-\ii \alpha J_{\rm z}^2/2J)}\exp{(-\ii \pi J_{\rm y}/2)}$.
We have chosen $\alpha=30$ ensuring fully chaotic corresponding classical dynamics,
with angular momentum coordinates on a unit sphere named as $x,y,z$, and 
$J=1000$ determining the effective Planck constant
$\hbar=1/J=10^{-3}$. 
The perturbation is chosen as $V=(J_{\rm x}^2-J_{\rm z}^2)/2J^2$ associated with
$W=J_{\rm z}^2/2J^2$. The initial 
state is either RIS (with Gaussian random expansion coefficients) 
or SU(2) CIS centered at $(\varphi,\theta)=(1,1)$. In both cases the
initial state is projected on an invariant subspace of dimension ${\cal N}=J$ 
spanned by 
${\cal H}_{\rm OE}=\{\ket{2m}-\ket{\!-\!2m},\ket{2m\!-\!1}+\ket{\!-\!(2m\!-\!1)}; 
m=1,\ldots,J/2\}$ where $\ket{m}$ is an eigenstate of $J_{\rm z}$. We first checked the 
plateau. Within the linear response (\ref{eq:LRP}) 
we have to evaluate only $\kappa^2_{\rm cl}=1/45$ for the corresponding classical observable 
$W_{\rm cl}=z^2/2$, giving $F^{\rm CIS}_{\rm plat}=1-(\delta J)^2 \kappa^2_{\rm cl}$, 
$F^{\rm RIS}_{\rm plat}=1-(\delta J)^2 2 \kappa^2_{\rm cl}$. These values give good agreement
with the fidelity for weak perturbation $\delta=10^{-3}$ shown in fig.~\ref{fig:1ktop}a, whereas
for strong perturbation $\delta=10^{-2}$ shown in fig.~\ref{fig:1ktop}b the theoretical values 
(\ref{eq:NLRP}) of $F_{\rm plat}$, 
expressed in terms of the generating function $G(z)$ for CIS/RIS, 
have to be calculated exactly, and indeed the
agreement is excellent. Integration over the sphere yields
$G(\delta J)=\sqrt{\frac{\pi}{2 \delta J}} {\rm erf}(e^{\ii \pi/4} \sqrt{\delta J/2}).$
Comparing with the asymptotic general formula for $G(z)$ (\ref{eq:Gasym})
we now also find a diffractive contribution due to oscillatory behavior of the complex erf-function.
Small (quantum) fluctuations around the theoretical plateau values in fig.~\ref{fig:1ktop}
lie beyond the leading order semiclassical description. 
In fig.~\ref{fig:1ktop} we also demonstrate that the semiclassical formula (\ref{eq:t2}) for $t_2$ 
works very well.
\begin{figure}
\centerline{\includegraphics[angle=-90,width=3.2in]{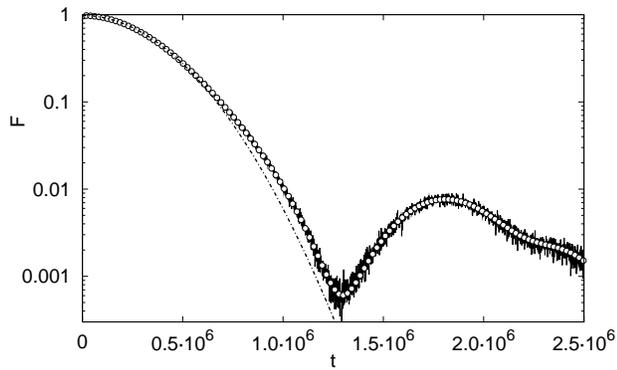}}
\caption{Long-time Gaussian decay for CIS of a single kicked top for the
same parameters as in 
fig.~\ref{fig:1ktop}a.
Full curve is a direct numerical evaluation, empty 
circles are numerical calculation using the renormalized strength $\delta_R$ and 
operator $R$, while the chain curve gives the theoretical decay (\ref{eq:Fgau}).
}
\label{fig:1ktopL}
\end{figure}
Long-time Gaussian decay for the parameters of fig.~\ref{fig:1ktop}a is shown
in fig.~\ref{fig:1ktopL}. Here we compare a direct numerical calculation with the numerical calculation 
using a renormalized perturbation strength $\delta_R$ and the effective perturbation operator 
(\ref{eq:defR}) 
$R=\frac{-1}{2J^3}(J_{\rm x}J_{\rm y}J_{\rm z}+J_{\rm z}J_{\rm y}J_{\rm x})$,
and with the theoretical prediction (\ref{eq:Fgau}) where the classical dynamics 
of $R_{\rm cl}=-xyz$ gives $\sigma=5.1\cdot 10^{-3}$.
\par
\begin{figure}
\centerline{\includegraphics[angle=-90,width=3.2in]{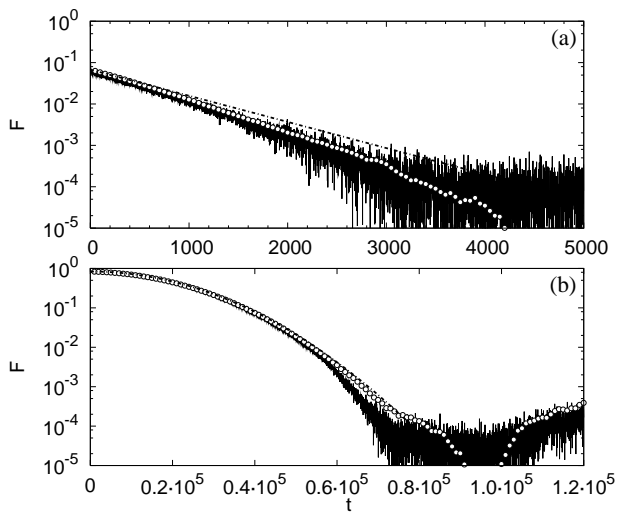}}
\caption{Long-time fidelity decay in two coupled kicked tops. For strong 
perturbation $\delta=7.5\cdot 10^{-2}$ (a) we obtain an exponential decay, and
for smaller $\delta=2\cdot 10^{-2}$ (b) we have a Gaussian decay. 
Meaning of the curves is the same as in fig.~\ref{fig:1ktopL}.
}
\label{fig:2ktop}
\end{figure}
To demonstrate the possibility of clean exponential 
long-time decay of fidelity (\ref{eq:Fexp}) we look at a system of two ($d=2$)
coupled tops $\vec{J}_1$ and $\vec{J}_2$ given by a propagator
$
U=\exp{(-\ii \varepsilon J_{\rm z1}J_{\rm z2})} \exp{(-\ii \pi J_{\rm y1}/2)}\exp{(-\ii \pi J_{\rm y2}/2)},
$
with the perturbation generated by $W=A_1 \otimes \mathbbm{1} + \mathbbm{1}\otimes A_2$, where 
$A=J_{\rm z}^2/2J^2$ for each top. We set $J=1/\hbar=100$, and $\varepsilon=20$ in order to be in a 
fully chaotic regime. The initial state is 
always a direct product of SU(2) coherent states centered at 
$(\varphi,\theta)=(1,1)$ which is subsequently projected on an invariant 
subspace of dimension ${\cal N}=J(J+1)$ spanned by 
$\{ {\cal H}_{\rm OE} \otimes {\cal H}_{\rm r} \}_{\rm sym}$, where 
${\cal H}_{\rm r}={\cal H} \setminus {\cal H}_{\rm OE}$ and $\{\cdot\}_{\rm sym}$ 
is a subspace symmetric with respect to the exchange of the two tops. 
The results of numerical simulation are shown in fig.~\ref{fig:2ktop}. Here 
we show only a long-time decay, as the situation in the 
plateau is qualitatively the same as for $d=1$. For large enough perturbation one 
obtains an exponential decay shown in fig.~\ref{fig:2ktop} (a), while for smaller 
perturbation we have a Gaussian decay shown in fig.~\ref{fig:2ktop} (b). 
Numerical data have been successfully compared with the theory
(\ref{eq:Fexp},\ref{eq:Fgau}) using classically calculated 
$\sigma=9.2 \cdot 10^{-3}$, and with the ``renormalized'' numerics using the operator 
$R$ (\ref{eq:defR}). 
\par
In this Letter we discussed a freeze of fidelity for arbitrary quantum
evolution provided only that the diagonal part of the perturbation in the basis of the
unperturbed evolution exactly vanishes. The value of the plateau can be arbitrary close to $1$ and
can span over arbitrary long time-ranges for sufficiently small strength of perturbation. We worked 
out in detail the case of systems with fully chaotic classical limit.
Our result is predicted to have immediate application to quantum information processing.
If combined with the inequality between the purity $I(t)$ of a reduced density matrix of
a bipartite quantum system and the fidelity, namely $I(t) > |F(t)|^2$ \cite{PSZ},
we predict that decoherence as characterized by $I(t)$ should also exhibit freeze 
for the particular class of perturbations.
Useful discussions with T.H.Seligman and financial support by the Ministry of Education, 
Science and Sports of Slovenia and DAAD19-02-1-0086, ARO United States are 
gratefully acknowledged.


\begin{thebibliography}{1}

\bibitem{fidelity}
A.~Peres, Phys. Rev. A {\bf 30}, 1610 (1984);
Ph.~Jacquod {\em et al.} Phys.~Rev.~E {\bf 64}, 
055203 (2001); N.~R.~Cerruti and S.~Tomsovic, Phys.~Rev.~Lett. {\bf 88}, 054103 (2002);
F.~M.~Cucchietti {\em et al.} Phys.~Rev.~E {\bf 65} 046209 (2002); 
G. Benenti and G. Casati, Phys.~Rev.~E {\bf 65} 066205 (2002).

\bibitem{jalabert}
R.~A.~Jalabert and H.~M.~Pastawski, Phys.~Rev.~Lett.~ {\bf 86}, 2490 (2001).

\bibitem{qcomp} 
M.~A.~Nielsen and I.~L.~Chuang, {\em Quantum computation and quantum information} 
(Cambridge Univ.~Press 2000).

\bibitem{VP} G.~Veble and T.~Prosen, Phys.~Rev.~Lett. {\em at press} (2004).

\bibitem{PZ} T.~Prosen and M.~\v Znidari\v c, J.~Phys.~A {\bf 35}, 1455 (2002); 
see also 
T.~Prosen and M.~\v Znidari\v c, J.~Phys.~A {\bf 34}, L681 (2001);
T.~Prosen, Phys. Rev. E {\bf 65}, 036208 (2002).

\bibitem{ifreeze} T.~Prosen and M.~\v Znidari\v c, New J. Phys. {\bf 5}, 109 (2003).

\bibitem{control} We note that the approximation (\ref{eq:BCH}) is accurate
for arbitrary long times, provided $\delta$ is sufficiently small, since
one can show directly that the third order ($\delta^3$) is again at most
proportional to $n$.

\bibitem{appr} In the following, $\approx$ means {\em equal in the leading order in} $\hbar$.

\bibitem{PSZ} T.~Prosen, T.~H.~Seligman and M.~\v Znidari\v c, Phys. Rev. A {\bf 67}, 
062108 (2003).


\end{thebibliography}
\end{document}